\documentclass[fleqn,twoside]{article}
\usepackage{espcrc2}
\usepackage{epsfig}
\usepackage{amssymb}
\usepackage{graphicx}
\usepackage{subfigure}

\newcommand{\mplank}{\bar{M}_{\rm  P}}

\def\be{\begin{equation}}
\def\ee{\end{equation}}
\def\bea{\begin{eqnarray}}
\def\eea{\end{eqnarray}}

\title{Quintessence  and the cosmological constant}
\author{M. Doran \address[HDTHEP]{Institut f\"ur Theoretische Physik,\\
   Philosophenweg 16,
   69120 Heidelberg, Germany} 
      and  C. Wetterich \addressmark[HDTHEP]}

\begin{document}
\begin{abstract}
Quintessence -- the energy density of a slowly
evolving scalar field -- may constitute a dynamical form 
of the homogeneous dark energy in the universe. We review
the basic idea
in the light of the cosmological constant problem. Cosmological
observations or a time variation of fundamental
`constants' can
distinguish quintessence from a cosmological constant.
\end{abstract}

\maketitle

The idea of quintessence originates from an attempt to understand
the smallness of the ``cosmological constant'' or dark energy in terms of the
large age of the universe \cite{Wetterich:1988fm}. As a characteristic consequence,
the amount of dark energy may be of the same order of magnitude as
radiation or dark matter during a long period of the cosmological 
history, including the present epoch. Today, the inhomogeneous energy
density in the universe -- dark and baryonic matter -- is about $\rho_{inhom}\approx(10^{-3} {\rm eV})^4$. This number is tiny in
units of the natural scale given by the Planck mass $M_p=1.22\cdot 10^{19}$ GeV. Nevertheless, it can be understood easily as a direct consequence 
of the long duration of the cosmological expansion: a dominant
radiation or matter 
energy density decreases $\rho\sim M^2_pt^{-2}$ and the
present age of the universe is huge, $t_0\approx 1.5\cdot 10^{10}$ yr.
It is a natural idea that the homogeneous part of the energy density 
in the universe -- the dark energy -- also decays with time and therefore
turns out to be small today\footnote{For some related ideas see ref.
\cite{CC}, \cite{SF}, \cite{Peccei:1987mm}.}.

A simple realization of this idea, motivated by the anomaly of the
dilatation symmetry, considers a scalar field $\phi$ with
an exponential potential \cite{Wetterich:1988fm}
\begin{equation}\label{1}
{\mathcal L}=\sqrt g\left\{\frac{1}{2}\partial^\mu\phi\partial_\mu\phi+
V(\phi)\right\}
\end{equation}
where
\begin{equation}
 V(\phi)=M^4\exp(-\alpha\phi/M),
\end{equation}
with $M^2=M^2_p/16\pi$. In the simplest version $\phi$ couples only to
gravity, not to baryons or leptons. Cosmology is then determined by the
coupled field equations for gravity and the scalar ``cosmon'' field
in presence of the energy density $\rho$ of radiation or matter. 
For a homogeneous and flat universe they read ($n=4$ for radiation and $n=3$ for nonrelativistic matter)
\begin{eqnarray}\label{AA}
&&H^2=\frac{1}{6M^2}\left(\rho+\frac{1}{2}\dot\phi^2+V\right),\nonumber\\
&&\ddot\phi+3H\dot\phi+\frac{\partial V}{\partial\phi}=0,\nonumber\\
&&\dot\rho+nH\rho=0.\end{eqnarray}
One finds that independently of the precise initial conditions the 
behavior for large $t$ approaches an exact ``cosmological attractor
solution'' (or ``tracker solution'') where the scalar kinetic and 
potential energy density scale proportional to  matter
or radiation \cite{Wetterich:1988fm}
\begin{eqnarray}\label{2}
&&\phi=\frac{2M}{\alpha}\ln(t/\bar t)\ ,\quad \frac{1}{2}\dot\phi^2
=\frac{2M^2}{\alpha^2}t^{-2},  \\
&& V=\frac{2M^2}{\alpha^2}
\frac{(6-n)}{n}t^{-2},
\end{eqnarray}
with the usual decrease of the Hubble parameter $H$
\begin{equation}\label{3}
H=\frac{2}{n}t^{-1}\quad,\quad \rho\sim t^{-2}.\end{equation}
This simple model predicts a fraction of dark energy or
homogenous quintessence (as
compared to the critical energy density $\rho_c=6M^2H^2$) which is
constant in time
\be\label{4}
\Omega_{\rm h}=\left(V+\frac{1}{2}\dot\phi^2\right)/\rho_c=
\rho_\phi/\rho_c=\frac{n}{2\alpha^2}\ee
both for the radiation-dominated $(n=4)$ and matter-dominated
$(n=3)$ universe $((\Omega_{\rm h}+\rho/\rho_c)=1)$.
This would lead to a natural explanation why 
today's dark energy is of the same order of magnitude as dark matter.

\begin{figure*}[!ht]
\begin{center}
\includegraphics[angle=-90,scale=0.5]{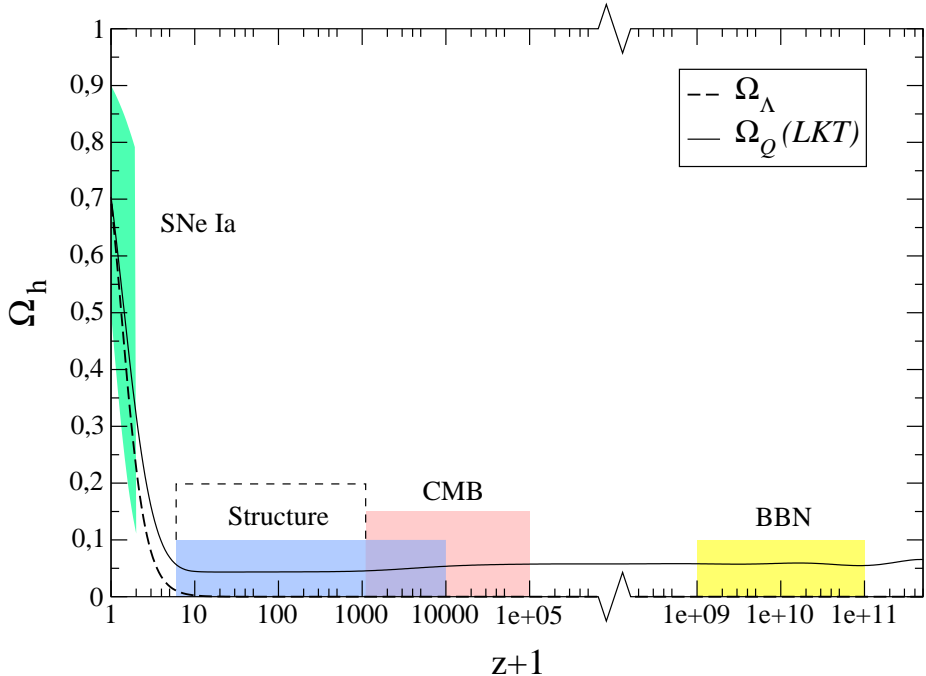}
\caption{Fractional dark energy contribution as a function of redshift $z$.
The shaded regions indicate bounds coming from Big Bang Nucleosynthesis (BBN),
the cosmic microwave background (CMB), structure formation and the
Supernovae Type Ia (SNe Ia) observations. The dashed curve shows a cosmological
constant with $\Omega_\Lambda^0  = 0.7$, the straight line a Leaping Kinetic
Term quintessence model with the same amount of dark energy today.
}\label{fig::dark}
\end{center}
\end{figure*}

The qualitative ingredients for the existence of the stable
attractor solution\footnote{For more details see refs.
\cite{rp,Frieman:1995pm,clw,wet}.} (\ref{2}), (\ref{3}) are
easily understood: for a large value of $ V(\phi)$ the force term
in eq. (\ref{AA}), $\partial V/\partial\phi=-(\alpha/M)V$, is large,
and the dark energy decreases faster than matter or radiation. In
the opposite, when the matter or radiation energy density is much 
larger than $V$, the force is small as compared to the damping term
$3H\dot\phi$ and the scalar ``sits and waits'' until the 
radiation or matter density is small enough such that the over-damped 
regime ends. Stability between the two extreme situations is reached
for $V\sim \rho$.

\begin{figure*}[!ht]
\begin{center}
\vspace*{.2cm}
\parbox[b]{15.7cm}{\psfig{width=15.5cm,file=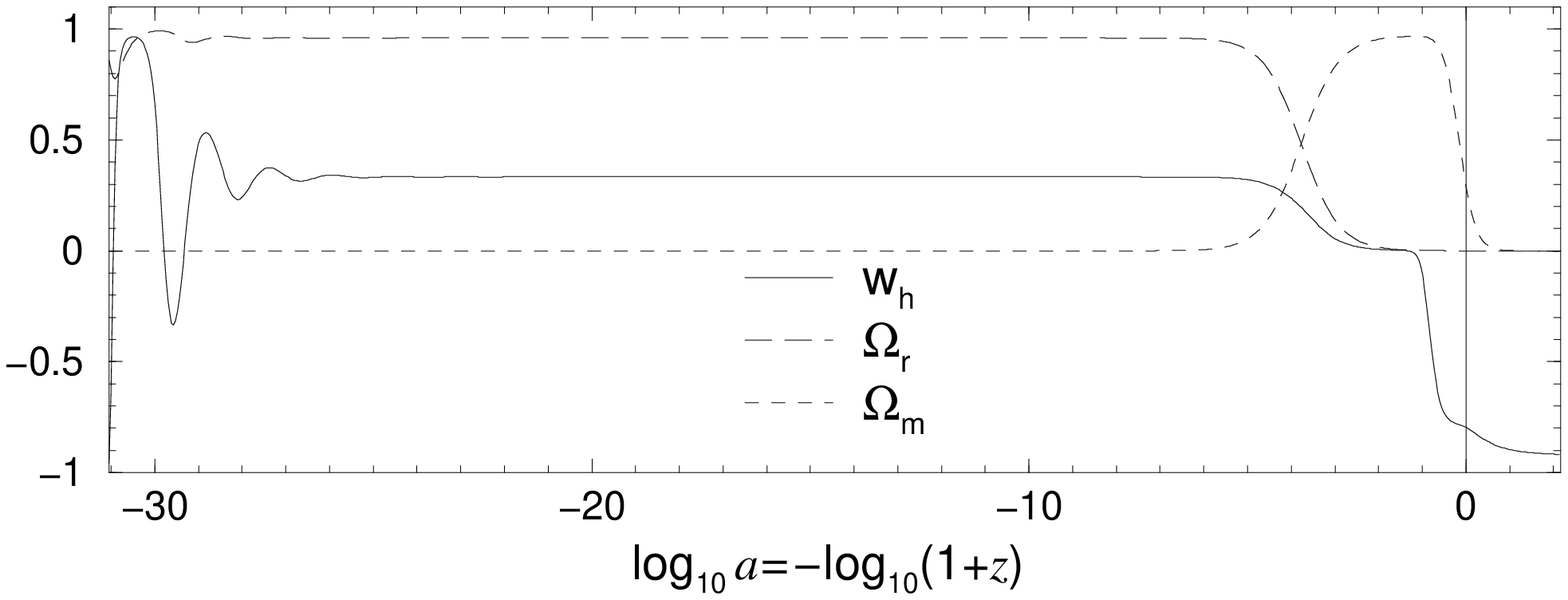}}\\
\end{center}
\refstepcounter{figure}\label{jump}
{\bf Figure \ref{jump}:} Cosmological evolution with a leaping kinetic term.
We show the fraction of energy in radiation ($\Omega_r$) and matter
($\Omega_m$) with $\Omega_{\rm h}=1-\Omega_{\rm r}-\Omega_{\rm m}$. The equation of state
of quintessence is specified by $w_{\rm h}$.
\end{figure*}

The scenario with constant $\Omega_{\rm h}$ is, however, too simple to be consistent
with observational data. In Fig. \ref{fig::dark}, we have collected our present knowledge
about the time evolution of $\Omega_{\rm h}$. 
From present observations one concludes that today's fraction of dark energy
is rather large
\be\label{5}
\Omega^0_{\rm h} =0.6-0.7.\ee
On the other hand, structure formation would be hindered by a too
large amount of dark energy \cite{fj}. One infers an approximate
upper bound for the average fraction of dark energy during structure formation \cite{JS}
\be\label{6}
\Omega^{\rm sf}_{\rm h} \lesssim 0.1 - 0.2.\ee
As a consequence, the fraction of dark energy must have
increased in the recent epoch since the formation of structure.
This implies a negative equation of state for quintessence \cite{cds},
\cite{swz} $w_{\rm h} =p_\varphi/\rho_\varphi<0$ and can lead to a universe
whose expansion is presently accelerating, as suggested by the 
redshifts of distant supernovae \cite{perl}.
For a phenomenological analysis of the observations that lead to the
constraints depicted in Fig. \ref{fig::dark}, 
we refer to \cite{Wetterich:2001jw,Bean:2002xy} and the 
other talks at this meeting.

The pure exponential potential in eq. (\ref{1}) cannot
 account for the recent increase in $\Omega_{\rm h}$. Possible modifications
of the basic idea of quintessence include the use of other potentials
\cite{Wetterich:1988fm,rp,Frieman:1995pm,as,bcn}, the coupling of quintessence to
dark matter \cite{wet}, \cite{ame}, nonstandard scalar kinetic terms
\cite{ams} or the role of nonlinear fluctuations \cite{CDM}. We note that
these ideas may not be unrelated, since the presence of large 
fluctuations can modify the effective field equations (e.g. change 
the effective cosmon potential and kinetic term) and lead to a coupling
between quintessence and dark matter \cite{CDM}.

In view of the still very incomplete theoretical understanding of the 
origin of quintessence the choice of an appropriate effective action for the
cosmon is mainly restricted by observation. For comparison with observation
and a discussion of naturalness of various approaches \cite{Heb} we
find it convenient to work with a rescaled cosmon field such that
 the scalar field Lagrangian reads
\be
{\mathcal L}(\varphi)=\frac{1}{2}\,(\partial\varphi)^2\,k^2(\varphi)+
\exp[-\varphi]\,.
\label{A}
\ee
Here and in what follows all quantities are measured in units of the reduced
Planck mass $\mplank$, i.e., we set $\mplank^2\equiv M_P^2/(8
\pi)\equiv(8\pi G_N)^{-1}=2M^2=1$. The Lagrangian of Eq.~(\ref{A}) contains
a simple exponential potential $V=\exp[-\varphi]$ and a non-standard kinetic
term with $k(\varphi)>0$.
The exponential potential in eq. (\ref{1}) corresponds to a constant $k=\frac{1}{\sqrt2\alpha}$.

This parameterization has the advantage of a one to one correspondence between 
the value of $\varphi$ and the contribution to dark energy from the cosmon potential $V$. 
At present, we have $\varphi \approx 276$ and the recent history of quintessence is 
directly connected to the behavior of $k(\varphi)$ for $\varphi$ smaller but in the
vicinity of this value. For many (`tracker') scenarios of quintessence, there is a direct relation between
the functions $k(\varphi)$ and $\Omega_{\rm h}(z)$, the latter being the one which is 
most easily accessible to observation and therefore the most appropriate one for phenomenological
discussions and comparison between different observations.

There are roughly two classes of quintessence models - one where the dark energy plays a role in the early
cosmology, i.e. at last scattering and before (early quintessence) and the other where it is 
negligible except for a relatively recent epoch (recent quintessence). 
Recent quintessence corresponds to a function $k(\varphi)$ that becomes tiny for $\varphi \ll 276$.
An example for early quintessence is given by leaping kinetic term (LKT) quintessence \cite{Heb},
where the effective value of $k$ jumps from a small to a larger value - possibly as
a back-reaction to structure formation or some other `recent' cosmological event.
\begin{eqnarray}
&&k(\varphi)=k_{min}+\mbox{tanh}(\varphi-\varphi_1)+1 \nonumber\\
&&(\mbox{with}\quad  
k_{min}=0.1\,,\,\,
\,\varphi_1=276.6\,)\,.\label{H}
\end{eqnarray}
The cosmological evolution of this model is depicted in Fig.~\ref{jump}. 
The LKT
model completely avoids the explicit use of very large or very small
parameters and  realizes all the desired features of quintessence \cite{Heb}. The 
homogeneous dark energy density tracks below the radiation or matter component in 
the early universe
($k=0.1$, corresponding to $\Omega_{\rm h} = 0.04$) and then suddenly comes to dominate the evolution when $k$ rises
to a value $k \gtrsim 1$. With a tuning on the percent level
(the value of $\varphi_1$ has to be appropriately adjusted) realistic
present-day values of $\Omega_{\rm h}^0$ and $w_{\rm h}^0$ can be realized. In the
above example, one finds $\Omega_{\rm h}^0=0.70$ and $w_{\rm h}^0=-0.80$.
Note that, due to the extended tracking period, the late cosmology is
completely insensitive to the initial conditions. 
For models of the type discussed in \cite{as}, $k(\varphi)$ diverges
near $\varphi_1$, leading to a similar cosmology.

In this talk, we turn back to the original question,
\cite{Wetterich:1988fm}, if quintessence can explain why the dark energy vanishes asymptotically
for large time, or, equivalently, if quintessence is related to a solution of the
cosmological constant problem. In terms of the cosmon potential this concerns
the question if the property $V(\varphi \to \infty) \to 0$ is natural.

We address this issue in the context of some (unknown) unified theory with a dynamical
unification scale $M_{\rm GUT}$. Similar to grand unified theories,
higher dimensional theories and superstring theories, the unification scale can be 
associated to the expectation value of a scalar field $\chi$. This field will play
the role of the cosmon. We restrict the discussion to the (four dimensional) system
of gravity and the cosmon with (euclidian action)
\begin{eqnarray}\label{eqn::action}
S &=& \int d^4x  \sqrt{\bar g} \Big [ - \frac{1}{12} f^2(\chi) \chi^2 \bar R \nonumber\\
&\ &\qquad + \frac{1}{2}Z_\chi(\chi) \partial_\mu \chi \partial_\nu \chi \bar g^{\mu\nu} + V(\chi) \Big ]
\end{eqnarray} 
The function $f(x)$ is slowly varying such that the dynamical Planck mass is 
essentially $\propto \chi$. The fate of the cosmological constant depends on the behavior of
the potential $V(\chi)$ for $\chi \to \infty$. Here, we only make the assumption that 
the combination $f^4 \chi^4 / V$ diverges for $\chi \to \infty$ and is a monotonic
function of $\chi$. Solving the field equations for this system \cite{Wetterich:1988fk},
one finds a stable solution where $\chi$ grows to infinity for large time. 

The interpretation of 
cosmology may be somewhat simpler, if we rescale the metric such that
the Planck mass becomes a constant, $\bar g_{\mu\nu} = (6 \mplank^2 / f^2 \chi^2) g_{\mu\nu}$.
The action (\ref{eqn::action}) becomes then precisely the one underlying our previous
discussion ($\mplank =1$)
\begin{eqnarray}
S &=& \int d^4x  \sqrt{g} \Big [ -\frac{1}{2} R + \frac{1}{2}k^2(\varphi) \partial_\mu \varphi  \partial^\mu \varphi  \nonumber\\
&\ &\qquad \qquad \qquad+ \exp(-\varphi) \Big ].
\end{eqnarray}
Here, $\varphi$ is related to $\chi$ by 
\begin{equation}
\varphi = \ln \frac{f^4(\chi) \chi^4}{36 V(\chi)}
\end{equation}
and
\begin{eqnarray}
k^2(\varphi) &=& \frac{3}{8}\left [ \frac{Z_\chi}{f^2} + \left(1 + \frac{\partial\ln f}{\partial \ln \chi}\right)^2 \right ] \nonumber\\
&\ & \times \left (1 + \frac{\partial \ln f}{\partial \ln \chi} - \frac{1}{4} \frac{\partial \ln V}{\partial \ln \chi} \right)^{-2} 
\end{eqnarray}
is a finite non-zero positive function. In particular, no additive constant is 
present in the cosmon potential.

We conclude that the cosmological constant vanishes independently of the 
details of $V(\chi)$ - for example we may add a constant in Equation (\ref{eqn::action}) -
provided the `cosmon condition' \cite{Wetterich:1988fm} 
\begin{equation}\label{eqn::beh}
\lim_{\chi \to \infty} \left(\frac{f^4 \chi^4}{V} \right) \to \infty
\end{equation}
holds\footnote{A similar situation arises if the limit is zero. Only
a finite non-zero limit would lead to a cosmological 
constant after Weyl scaling.}.
The theoretical status of the cosmon condition cannot be settled unless we know
the properties of the unified theory. At least, the asymptotic vanishing of 
the cosmological constant is associated to some generic behavior (\ref{eqn::beh}) and
needs no tuning of parameters to many decimal places. The status of
naturalness of the cosmon condition is linked to the role of quantum 
fluctuations. Here, we point out only two observations:\\
(i) The quantum effects on the potential $V(\chi)$ are dominated by
fluctuations with momenta around the unification scale $\chi$. Fluctuations 
of the fields of the standard model, the cosmon\footnote{The exponential
potential is actually stable with respect to the cosmon fluctuations \cite{Doran:2002bc} },
or graviton
with momenta much
smaller than $\chi$ play a negligible role. The fluctuations of the modes relevant
around the unification scale may be described by a higher dimensional 
theory or string theory such that simple extrapolations from four dimensional quantum
fluctuations may be quite inappropriate. Arguments that arbitrarily select parts of
the fluctuation effects (like quantum fluctuations of the standard model fields with
momenta below a fixed scale) and associate their size with the `natural size' of the
total fluctuation effect are known to be grossly misleading in other circumstances, like
critical behavior in statistical physics. For example, the
`quartic coupling' $\lambda \propto V/\chi^4$ may have a `renormalization group flow'
to zero for $\chi \to \infty$ and therefore become very small at large $\chi$, whereas 
arbitrarily selected `individual contributions' of quantum fluctuations are much
larger.\\[1ex]
(ii) The naive guess that the quantum corrections to $V$ should be $\propto \chi^4$
by dimensional reasons tacitly assumes that dilatation symmetry is a quantum
symmetry. This is, however, not very likely and one expects a dilatation anomaly.
If this manifests itself in the form of a non-vanishing anomalous dimension A for $\lambda$,
one rather has $V \propto \chi^{4-A}$.
\\[1ex]

As an important consequence of the dilatation anomaly, also the
gauge couplings and the  mass ratio between the nucleon and the Planck mass 
typically become
dependent on $\chi$. An  example 
for varying gauge couplings is provided
by a term (with $F_{\mu\nu}$ the field
strength of the gauge fields)
\begin{equation}
\mathcal{L}_{\rm F} = \frac{1}{4} Z_{\rm F}(\chi) F^{\mu\nu}F_{\mu\nu},
\end{equation}
with anomalous dimension 
\begin{equation}
\eta_F = \frac{\partial\ln Z_F}{\partial\ln\chi}.
\end{equation}
Due to the cosmological variation of $\chi$ this leads to a time dependence of
couplings \cite{Wetterich:1988fm,wet,Dvali:2002dd,Chiba:2001er,Wetterich:2002ic}
and a composition dependent gravity-like long range force
 \cite{Peccei:1987mm,wet,Dvali:2002dd,Chiba:2001er,Wetterich:2002ic}.
The time variation of fundamental couplings is a generic prediction of quintessence,
albeit the size of the effect is not known. The recently reported 
cosmological time variation of the fine structure constant \cite{Webb:2001mn}
corresponds to $\eta_F = -4\times 10^{-6}A$. Confirmation
of this result would 
be a very clear signal for quintessence - in distinction to the
cosmological constant for which no such variation is expected.\\[1.5ex]

\noindent{\bf Acknowledgment:} The authors would like to thank
A. Hebecker, J. J\"ackel, M. Lilley, and M. Schwindt for collaboration
on the content of this work. Part of it is based on refs. \cite{Heb}, \cite{JS}.


\begin{thebibliography}{99}

\bibitem{Wetterich:1988fm}
C.~Wetterich,
Nucl.\ Phys.\ B {\bf 302} (1988) 668.

\bibitem{CC}   M. \"Ozer and M.O. Taha, Phys. Lett. B171 (1986) 363;\\
                K. Freese, F.C. Adams, J.A. Frieman and E. Mottola, Nucl.
                Phys. B287 (1987) 797;\\
                M. Reuter and C. Wetterich, Phys. Lett. B188 (1987) 38


\bibitem{SF}   A.D. Dolgov, in {\it The Very Early Universe: Proc. of the
                1982 Nuffield Workshop at Cambridge}, ed. by G.W. Gibbons,
                S.W. Hawking and S.T.C. Siklos (Cambridge Univ. Press)
                p. 449;\\
                L.F. Abbott, Phys. Lett. B150 (1985) 427;\\
                T. Banks, Nucl. Phys. B249 (1985) 332;\\ 
                S.M. Barr, Phys. Rev. D36 (1987) 1691

\bibitem{Peccei:1987mm}
R.~D.~Peccei, J.~Sola and C.~Wetterich,
Phys.\ Lett.\ B {\bf 195} (1987) 183.

\bibitem{rp}    B. Ratra and P.J.E. Peebles,  Astrophys. J. Lett. 325
                (1988) L17, Phys. Rev. D37 (1988) 3406


\bibitem{Frieman:1995pm}
J.~A.~Frieman, C.~T.~Hill, A.~Stebbins and I.~Waga,
Phys.\ Rev.\ Lett.\  {\bf 75} (1995) 2077
[arXiv:astro-ph/9505060].



\bibitem{clw}   E. J. Copeland, A.R. Liddle and D. Wands, Ann. N. Y. Acad.
                Sci. 688 (1993) 647, Phys. Rev. D57 (1998) 4686

\bibitem{wet}   C. Wetterich, Astron. Astrophys. 
                301 (1995) 321 (hep-ph/9408025)


\bibitem{fj}    P.G. Ferreira and M. Joyce, Phys. Rev. Lett. 79 (1997) 4740,
                Phys. Rev. D58 (1998) 023503



\bibitem{JS}
M.~Doran, J.~M.~Schwindt and C.~Wetterich,
Phys.\ Rev.\ D {\bf 64} (2001) 123520
[astro-ph/0107525].



\bibitem{cds}   R.R. Caldwell, R. Dave and P.J. Steinhardt, Phys. Rev.
                Lett. 80 (1998) 1582

\bibitem{swz}   P.J. Steinhardt, L. Wang and I. Zlatev, Phys. Rev. Lett. 82
                (1999) 896, Phys. Rev. D59 (1999) 123504

\bibitem{perl}  S. Perlmutter et al., Astrophys. J. 517 (1998) 565;\\
                A.G. Riess et al., Astron. J. 116 (1998) 1009

\bibitem{Wetterich:2001jw}
C.~Wetterich,
arXiv:astro-ph/0110211.

\bibitem{Bean:2002xy}
R.~Bean and A.~Melchiorri,
Phys.\ Rev.\ D {\bf 65} (2002) 041302
[arXiv:astro-ph/0110472].



\bibitem{as}    A. Albrecht and C. Skordis, Phys. Rev. Lett. 84 (2000) 2076

\bibitem{bcn}   T. Barreiro, E.J. Copeland and N.J. Nunes, Phys. Rev. D61
                (2000) 127301

\bibitem{ame}   L. Amendola, Phys. Rev. D62 (2000) 043511;\\
                R. Bean and J. Magueijo, astro-ph/0007199




\bibitem{ams}   C. Armendariz-Picon, V. Mukhanov and P.J. Steinhardt,
                astro-ph/0004134 and astro-ph/0006373

\bibitem{CDM}   C. Wetterich, hep-ph/0108266; astro-ph/0108411

\bibitem{Heb}    A. Hebecker, C. Wetterich, Phys. Lett. {\bf B497} (2001)
                281



\bibitem{Wetterich:1988fk}
C.~Wetterich,
Nucl.\ Phys.\ B {\bf 302} (1988) 645.

\bibitem{Doran:2002bc}
M.~Doran and J.~Jaeckel,
arXiv:astro-ph/0203018.

\bibitem{Dvali:2002dd}
G.~R.~Dvali and M.~Zaldarriaga,
Phys.\ Rev.\ Lett.\  {\bf 88} (2002) 091303
[arXiv:hep-ph/0108217].

\bibitem{Chiba:2001er}
T.~Chiba and K.~Kohri,
arXiv:hep-ph/0111086.


\bibitem{Wetterich:2002ic}
C.~Wetterich,
arXiv:hep-ph/0203266.



\bibitem{Webb:2001mn}
J.~K.~Webb {\it et al.},
Phys.\ Rev.\ Lett.\  {\bf 87} (2001) 091301
[arXiv:astro-ph/0012539].

\end{thebibliography}
\end{document}